\documentclass[twocolumn,showpacs,preprintnumbers,amsmath,amssymb]{revtex4}

\usepackage{graphicx}
\usepackage{dcolumn}
\usepackage{bm}

\begin{document}


\title{Stripe charge ordering in SrO-terminated SrTiO$_3$(001) surfaces}

\author{Katsuya Iwaya,$^{1}$ Ryota Shimizu,$^{1}$ Takeo Ohsawa,$^{1}$ Tomihiro Hashizume,$^{1,2,3}$ and Taro Hitosugi,$^{1}$}
\affiliation{
$^{1}$WPI-Advanced Institute for Materials Research, Tohoku University, Sendai 980-8577, Japan \\
$^{2}$Advanced Research Laboratory, Hitachi, Ltd., Saitama 350-0395, Japan \\
$^{3}$Department of Physics, Tokyo Institute of Technology, Tokyo 152-8551, Japan \\
}

\date{\today}

\begin{abstract}
The local electronic structure of the SrO-terminated SrTiO$_3$(001) surface was explored 
using scanning tunneling microscopy. At low bias voltages in the empty states, 
a unidirectional structure with a periodicity of 3 unit cells, superimposed on a $c(2\times2)$ reconstructed structure, 
was found to develop along the crystallographic $a$ axis. 
This structure indicates a charge-ordered stripe induced by carrier doping from oxygen vacancies in the SrO and the subsurface TiO$_2$ planes. 
In the filled states, localized deep in-gap states were observed in addition to large energy gaps in the tunneling 
spectra. This result represents inelastic tunneling due to significant electron-lattice interaction associated with unidirectional lattice distortion in the SrO-terminated surface. 
\end{abstract}


\maketitle

\section{INTRODUCTION}
Atomically controlled growth of transition metal oxide heterostructures has opened up 
new perspectives for surface and interface physics \cite{Schlom}, which originate from the interplay between spin, charge 
and orbital degrees of freedom \cite{Tokura}. Among the various substrates, strontium titanate SrTiO$_3$ is the most commonly used 
for growing perovskite heterostructures. SrTiO$_3$ has two possible (001) surface terminations, 
either a SrO or TiO$_2$ layer, which determine the stacking sequence of the deposited films, 
and thus control the electronic properties of the heterostructures \cite{Kawasaki,Bachelet}.   
To further explore unique functionalities at these surfaces and interfaces, 
an understanding of the electronic states of these two surface terminations at atomic resolution is highly desirable. 
In the TiO$_2$-terminated surface, numerous surface reconstructions that are dependent on the surface preparation conditions 
have been reported in single crystals \cite{Castell,Kubo} and thin films \cite{Iwaya_APEX,Ohsawa},  
whereas the SrO-terminated surface has remained relatively unexplored \cite{Jiang}. 

It is widely known that SrTiO$_3$ prepared in highly reduced conditions exhibits a
metallic behavior due to electron doping from oxygen vacancies \cite{Ohtomo}. 
Considering the very high dielectric constant of SrTiO$_3$ \cite{Muller}, a strong coupling between electrons and ions can be expected and is likely to play a crucial role in transport properties, as observed in various systems 
\cite{Stoneham}.   
The transport properties of SrTiO$_3$ have been understood in terms of large polarons, according to optical measurements \cite{Mechelen}. 
Furthermore, recent optical and angle resolved photoemission (ARPES) studies found a moderate electron-phonon 
coupling and no clear signature of small polarons in SrTiO$_3$ \cite{Mechelen,Meevasana}.  
ARPES is a surface-sensitive technique that can distinguish whether the electronic structure originates
in a bulk state or surface state. The electron-phonon interaction discussed by Meevasana {\it et al.} was attributed mainly 
to the bulk electronic structure rather than the surface state \cite{Meevasana}. 
Electron-lattice interactions near the surface could have a significant impact on the electronic properties 
at the SrTiO$_3$-based interface, however, an understanding of such surface-related effects has remained elusive.

In this paper, we have investigated the local electronic structure of the SrO-terminated SrTiO$_3$(001) surface using low-temperature scanning tunneling microscopy and spectroscopy (STM and STS). 
A stripe-like unidirectional ordering is identified at low energies in the empty states, 
which can be attributed to ordering of the Ti$^{3+}$ ions. 
In the tunneling spectra, large energy gaps with no density of states (DOS) at the Fermi level ($E_F$) 
are observed. These results suggest the importance of the electron-lattice interaction associated with the unidirectional lattice distortion induced by charge ordering in the SrO-terminated SrTiO$_3$ surface.

\section{METHODS}
Nb-doped (0.1 at.\%) SrTiO$_3$(001) single crystals and 10 unit cell (u.c.) thick SrTiO$_3$(001) thin films were examined. 
The single crystal surfaces were prepared by buffered HF etching \cite{Kawasaki} and subsequent annealing at 800$^{\rm{o}}$C for 6 h in ultrahigh vacuum (UHV), followed by 1000$^{\rm{o}}$C annealing for 3 min in UHV conditions.
The SrTiO$_3$(001) thin films were homoepitaxially grown on Nb-doped (0.1 at.\%) SrTiO$_3$(001) substrates at 1100$^{\rm{o}}$C in an oxygen pressure of $\sim 1\times10^{-5}$ Torr 
by pulsed laser deposition. Details regarding the film growth were reported previously \cite{Iwaya_APEX,Ohsawa}. 
In both samples, SrO layers with almost identical features were observed in STM images, in contrast to the TiO$_2$-terminated surfaces,
which exhibited completely different atomic arrangements depending on the surface preparation condition 
\cite{Castell,Kubo,Iwaya_APEX, Ohsawa}. 
The samples were transferred to an UHV STM system (Unisoku USM-1300S) without exposure of the surface to air, 
and all STM measurements were performed at 78 K.

\section{RESULTS AND DISCUSSION}
Figure 1(a) displays a wide-scale STM image of a SrTiO$_3$(001) single crystal 
that has a typical TiO$_2$-terminated step-and-terrace structure with a single unit cell step height ($\sim 0.4$ nm), 
as indicated in Fig.~1(b). Additionally, small terrace structures of intermediate height relative to the surrounding TiO$_2$-terminated surfaces were identified [Fig.~1(b)]. 
Considering the crystal structure and the height difference, it is reasonable to suppose that the small terrace is related 
to the SrO plane. In fact, this is consistent with previous friction force microscopy and coaxial-impact-collision ion-scattering spectroscopy 
(CAICISS) measurements, which identified two different surface terminations and that the minority fraction of the surface 
was a SrO layer with a half unit cell step height \cite{Iwahori}.

\begin{figure}
\includegraphics[width=8.4cm, bb=0 0 289 136]{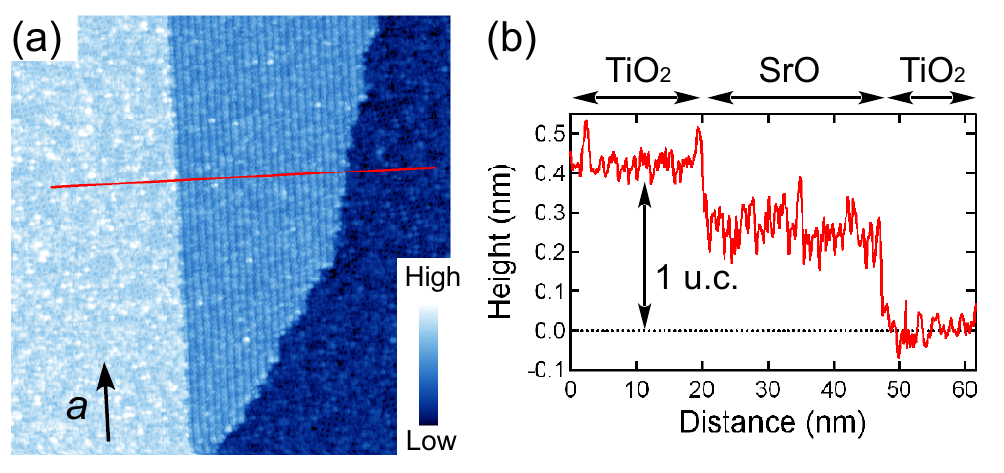}
\caption{(color online)
(a) Constant-current STM image of SrTiO$_3$(001) single crystal. ($70\times70$ nm$^2$. Sample-bias-voltage $V_s = +2.4$ V, and the set-point tunneling current $I_t = 30$ pA).
(b) Line profile across the steps in (a).    
}
\label{fig_1}
\end{figure}

Focusing at a finer scale, the atomic-scale structure of the SrO-terminated surface was identified. 
Figure 2(a) shows an STM image at a sample bias voltage ($V_s$) of +1.6 V, 
in which we observed a stripe-like unidirectional structure along the $a$ axis with a periodicity of 3 u.c.\ 
and a complex superimposed structure. To reveal this structure more clearly, 
the 3 u.c.\ spots were filtered out in the fast Fourier transformed (FFT) image [Fig.~2(b)], 
and an inverse-FFT image was obtained [Fig.~2(c)]. The filtered image clearly shows a $c(2\times2)$ structure 
with an inhomogeneous background. The $c(2\times2)$ structure is possibly due to alternate buckling of the TiO$_6$ octahedra. 
The $c(2\times2)$ structure has been observed in SrO-terminated surfaces of other perovskite systems, 
such as Sr$_2$RuO$_4$ \cite{Matzdorf} and Sr$_3$Ru$_2$O$_7$ \cite{Iwaya_PRL}. 
In this study, however, the $c(2\times2)$ superstructure coexists with the unidirectional order 
and with a periodicity of 3 u.c.  
The unit cell of this superstructure is therefore $c(6\times2)$, 
which is consistent with previous STM and low-energy electron diffraction (LEED) measurements \cite{Jiang,Shiraki}.   
In those measurements, the $c(6\times2)$ structure was assumed to be a surface reconstruction. 
However, the 3 u.c.\ modulation is aligned on every other unit of the $c(2\times2)$ lattice, 
as illustrated in Figs.~2(a) and (d). This indicates the possibility of electronic origin, rather than 
the normal surface reconstruction.

\begin{figure}
\includegraphics[width=8cm, bb= 0 0 256 289]{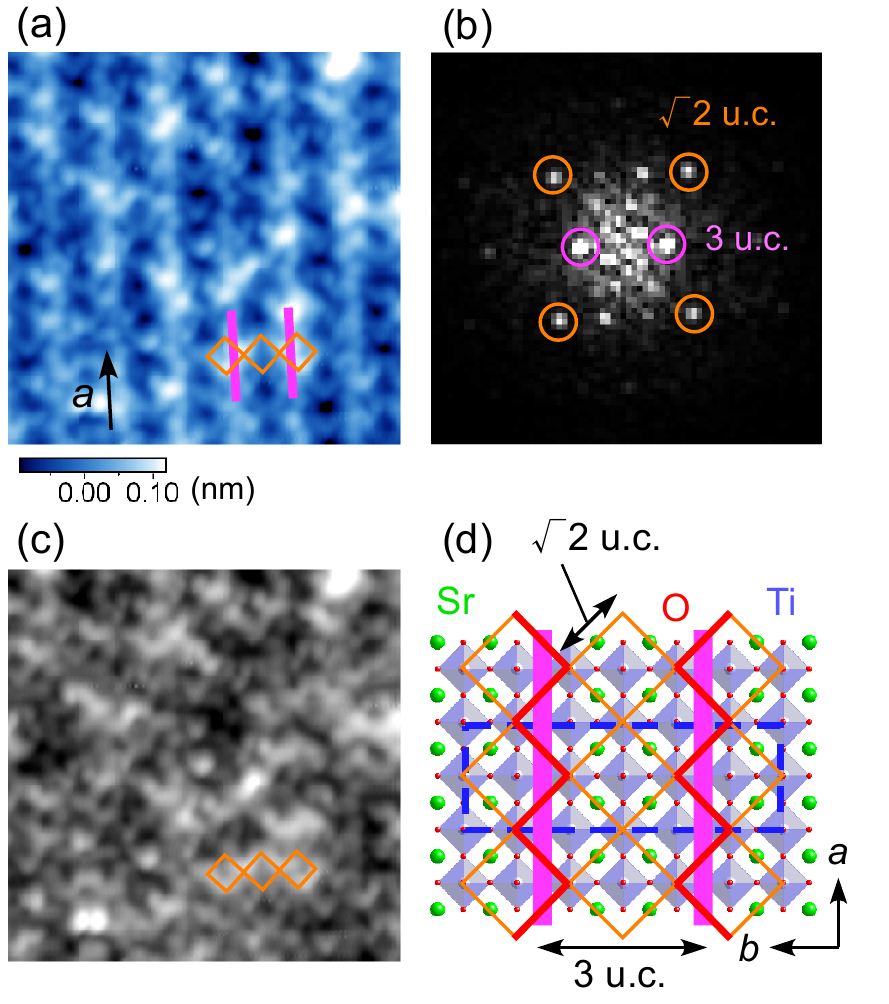}
\caption{(color online)
(a) STM image of a SrO-terminated SrTiO$_3$(001) single crystal surface at $V_s = +1.6$ V. 
$I_t = 30$ pA. $15 \times 15$ nm$^2$.
The 3 u.c.\ unidirectional structure and the $c(2\times2)$ structure are shown as lines and squares, respectively.
(b) FFT image corresponding to (a).
(c) Filtered STM image of (a). The 3 u.c.\ spots in (b) are filtered out and the $c(2\times2)$ structure is indicated as squares.
(d) Crystal structure of SrTiO$_3$(001) showing both the 1st (SrO plane) and 2nd layers (TiO$_2$ plane). 
The $c(2\times2)$ structure, 3 u.c.\ unidirectional structure, and the zig-zag structure are illustrated by the thin, thick and medium lines, respectively. 
The $c$($6\times2$) unit cell is shown by a dotted line.
}
\label{fig_2}
\end{figure}

Indeed, we observed a striking bias-voltage dependence of the STM image, as shown in Figs.~3(a--c). 
The unidirectional structure and the $c(6\times2)$ structure were clearly identified at $+1.7$ V [Fig.~3(a)], 
and at higher $V_s$ the unidirectional feature looked more enhanced [Fig.~3(b)].  
At $V_s > +3.3$ V, the unidirectional structure almost disappeared
and highly defective zig-zag chains were observed along the $a$ axis [Fig.~3(c)]. 
To demonstrate the evolution of the unidirectional structure more clearly, 
autocorrelation analyses of STM images were performed at each bias voltage. 
Figure 3(d) shows the intensity at 3 u.c.\ modulation in the autocorrelation image 
as a function of bias voltage. 
At low bias voltages, the intensity started to increase and was most enhanced at $\sim +2.1$ V. 
However, at $V_s > +3.3$ V, the intensity was almost negligible, which indicates the absence of unidirectional ordering. 
This strong bias dependence strongly supports the electronic origin of the unidirectional structure.  

\begin{figure}
\includegraphics[width=8.4cm, bb=0 0 251 231 ]{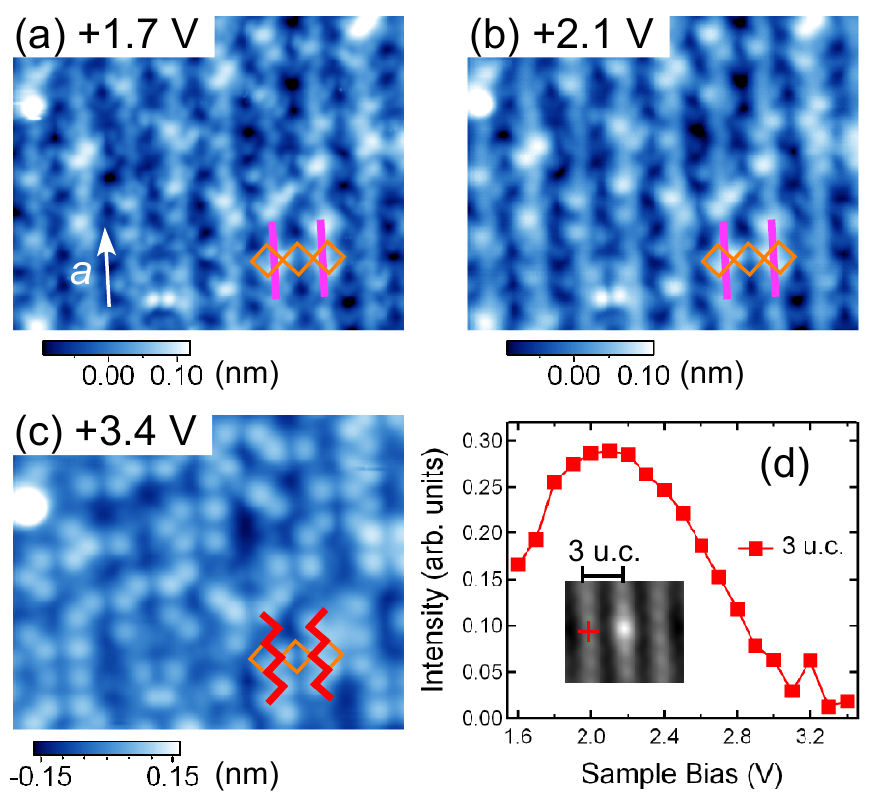}
\caption{(color online)
(a) STM image of a SrO-terminated SrTiO$_3$(001) single crystal surface at $V_s = +1.7$ V.
$I_t = 30$ pA. $15 \times 12$ nm$^2$. 
The 3 u.c.\ unidirectional structure and the $c(2\times2)$ structure are shown in the same way as in Fig.~2(d). 
(b--c) STM image at $V_s = +2.1$ V, $+3.4$ V, respectively. These were taken at the same location as (a).
The 3 u.c.\ unidirectional structure, the $c(2\times2)$ structure and the zig-zag structure are indicated 
in the same way as in Fig.~2(d).
(d) Plot of the autocorrelation intensity of the 3 u.c.\ periodicity as a function of bias voltage.
The autocorrelation image of (a) is shown in the inset. The location corresponding to the 3 u.c.\ periodicity is indicated by a cross.
}
\label{fig_3}
\end{figure}

The electronic states responsible for the tunneling originate mostly from the orbital states of subsurface Ti atoms, 
even in the SrO-terminated surface because the weakly hybridized Sr$^{2+}$ ions contribute a small amount of spectral weight to 
the DOS at the bias voltages investigated in this study \cite{Heifets}. 
Therefore, we can expect that the tunneling current at low bias voltages probes mainly the electronic states of the subsurface TiO$_2$ layer, 
whereas the electronic states of the SrO layer (mostly O) predominate at high bias voltages. 
Given these results, the unidirectional structure is likely to be induced in the TiO$_2$ subsurface layer, 
rather than in the SrO layer. 

One plausible explanation for the observed unidirectional modulation is a charge density wave (CDW), 
which has been encountered in a wide variety of layered materials, 
and is basically driven by Fermi surface nesting. 
A recent ARPES measurement revealed that the bulk-related Fermi surface topography of electron-doped SrTiO$_3$ 
consists of three ellipsoid-like surfaces along each of the three axes, 
while a cylinder-like Fermi surface along $k_z$ has been observed for the surface states \cite{Meevasana}. 
This result excludes the possibility of the unidirectional Fermi surface nesting required for a stripe-like ordering.
  
Another explanation would be charge ordering. 
A similar unidirectional 3 u.c.\ modulation, attributed to stripe charge ordering, 
has been observed in the manganite perovskite system La$_{1-x}$Ca$_x$MnO$_3$ ($x =0.67$) \cite{Mori}. 
We can expect an analogous stripe ordering composed of an array of Ti$^{3+}$ ions, 
which have one $t_{2g}$ electron on each site ($d^1$ configuration). 
Assuming that Ti$^{3+}$ ions form zig-zag chains to develop the stripe structure, 
as illustrated in Figs.~2(d) and 3(c), the density of doped electrons required for this ordering is $\sim 2\times10^{14}$ cm$^{-2}$. 
The dominant dopants are most likely oxygen vacancies because the surfaces were prepared under highly reducing conditions. Thus, we will now discuss the density of oxygen vacancies.
 
The bright spots consisting of a highly defective zig-zag structure at high bias voltages [Fig.~3(c)] can be attributed to oxygen vacancies in the SrO plane. 
In the rutile TiO$_2$(110) surface, oxygen vacancies were imaged as bright features in the empty states, 
due to the presence of dangling bonds at the Ti sites \cite{Minato}. 
Assuming a similar situation for the oxygen vacancies in the SrO surface,
the density of bright spots (oxygen vacancies) in the SrO plane
was estimated to be $\sim 5.3\times10^{13}$ cm$^{-2}$.

In order to measure the density of oxygen vacancies in the subsurface TiO$_2$ layer, 
STS (conductance map) measurements were performed from +3.5 to 0 V with a spatial resolution 
of $128\times128$ pixels in a $25\times25$ nm$^2$ region. 
The STM image taken simultaneously during the measurement is shown in Fig.~4(a). 
Many bright spots with an almost random distribution were identified in the conductance map at +3 V [Fig.~4(b)] 
with a characteristic peak or a shoulder structure in d$I$/d$V$ spectra [Fig.~4(c)]. 
We note that the peak and the shoulder structure at $V_s \sim +3$ V were much larger than the DOS 
in the range of $V_s < \sim +2.5$ V and therefore d$I$/d$V$ curves in Fig.~4(c) looked fully gapped. 
The inset of Fig.~4(c) shows a typical d$I$/d$V$ spectrum just above $E_F$, and the $n$-type behavior was clearly identified.  

Bright spots with similar spectroscopic features have also been observed in the TiO$_2$-terminated surfaces of SrTiO$_3$(001) substrates \cite{Ohsawa} 
and thin films \cite{Iwaya_APEX,Ohsawa,Iwaya}, 
which are considered to be due to oxygen vacancies based on the oxygen pressure dependence during the surface preparation \cite{Iwaya}. 
The different spectroscopic feature (the peak or the shoulder structure) is possibly due to the different site of 
the oxygen vacancy and might be associated with different charge densities,   
as observed in the TiO$_2$-terminated surfaces as well \cite{Iwaya_APEX,Ohsawa}.  
The density of bright spots estimated from the conductance map was $\sim 3.3\times10^{13}$ cm$^{-2}$, 
which is in the same order as that for the TiO$_2$-terminated surface \cite{Iwaya_APEX,Ohsawa}. 
In addition, the locations of bright spots did not correspond to the sites of the zig-zag structure; 
therefore, most of the bright spots observed in the conductance maps at +3 V originate in oxygen vacancy defects 
in the underlying TiO$_2$ layer rather than in the SrO layer.

\begin{figure}
\includegraphics[width=8cm, bb=0 0 280 333]{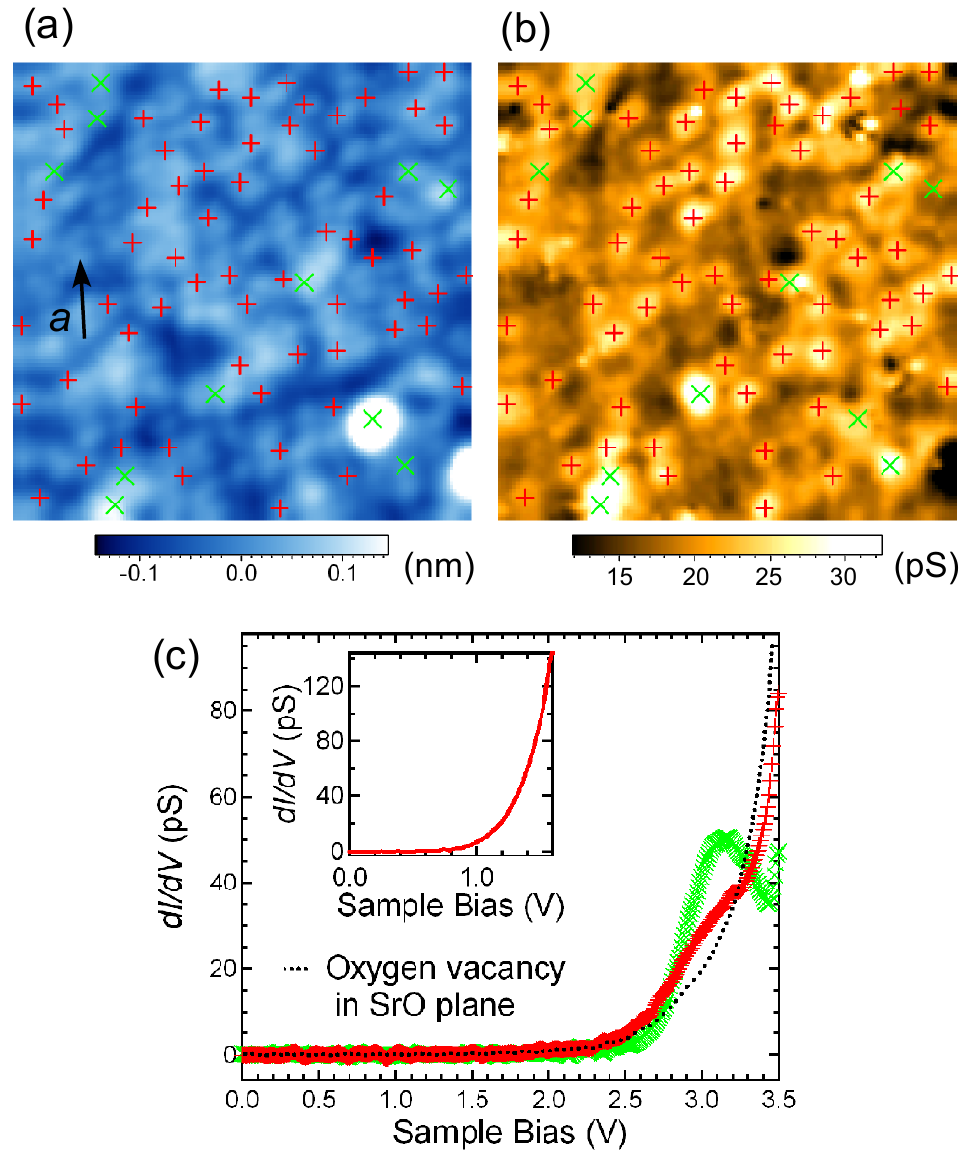}
\caption{(color online)
(a) STM image of SrO-terminated SrTiO$_3$(001) single crystal surface simultaneously acquired during STS measurement. 
($15 \times 15$ nm$^2$, $V_s = +3.5$ V, $I_t = 30$ pA).  
(b) Conductance map at $V_s = +3$ V measured at the same location as (a). 
The sites of oxygen vacancies in the subsurface TiO$_2$ layer are indicated by $+$ and $\times$.  
(c) d$I$/d$V$ spectra measured at oxygen vacancies in the subsurface TiO$_2$ plane. The same symbols as those in (a) and (b) 
are used to show the locations where the spectra were measured. 
For comparison, the d$I$/d$V$ spectrum at an oxygen vacancy in the SrO plane is also shown. Typical d$I$/d$V$ spectrum 
just above $E_F$ is shown in the inset. 
}
\label{fig_4}
\end{figure}

Since a neutral oxygen vacancy will donate two electrons, 
the total amount of electrons doped from oxygen vacancies in the SrO and the subsurface TiO$_2$ planes is estimated to be
 $(5.3\times10^{13} + 3.3\times10^{13}) \times 2 = 1.7\times10^{14}$ cm$^{-2}$, 
which is in good agreement with the expected value ($\sim 2.0\times10^{14}$ cm$^{-2}$). 

In the case of La$_{1-x}$Ca$_x$MnO$_3$, the stripe structure was mainly due to a strain induced by Jahn--Teller distortion of the MnO$_6$ octahedra \cite{Mori}. 
Notably, the unidirectional structures observed in this study were always found to be aligned along the step edge of the neighboring upper TiO$_2$ terrace, 
i.e., along the $a$ or $b$ axis, depending on the step edge structure [Fig.~1(a)]. 
Therefore, a substantial lattice distortion is possibly induced by the step edge as reported previously \cite{Jiang} 
and would be responsible for the unidirectional ordering.
 
To probe the electronic states in the filled states, STS measurements were performed from +1.5 to $-3$ V, 
because it was extremely difficult to obtain stable STM images in the filled states due to a large energy gap. 
Figure 5(a) shows the STM image of the SrO-terminated thin film surface at $V_s = +1.5$ V, simultaneously acquired during the STS measurement, 
exhibiting the stripe structure. We note that essentially the same results were obtained both in single crystals 
and thin films. A typical d$I$/d$V$ spectrum of the stripe had no DOS around $E_F$ and in the filled states up to $-3$ V [Fig.~5(b)]. 
However, the conductance map in the range of $V_s < -2$ V had quite a few bright spots [Fig.~5(c)], 
accompanied by a broad peak structure at $\sim -2.2$ V in the d$I$/d$V$ spectrum, 
as shown in Fig.~5(b). The d$I$/d$V$ spectra far away from the bright spots indicated a large energy gap of $\sim 3.7$ V, 
which was slightly larger than the bulk band gap, 
and the energy gap was almost independent of the surface termination [Fig.~5(d)]. 
This result is in contrast to a naive assumption in which a smaller band gap is expected in the TiO$_2$-terminated surface 
than in the SrO-terminated surface, due to a shift of the valence band top caused by weaker hybridization between Ti and O atoms \cite{Heifets_2}. 

\begin{figure}
\includegraphics[width=8.4cm, bb=0 0 286 205]{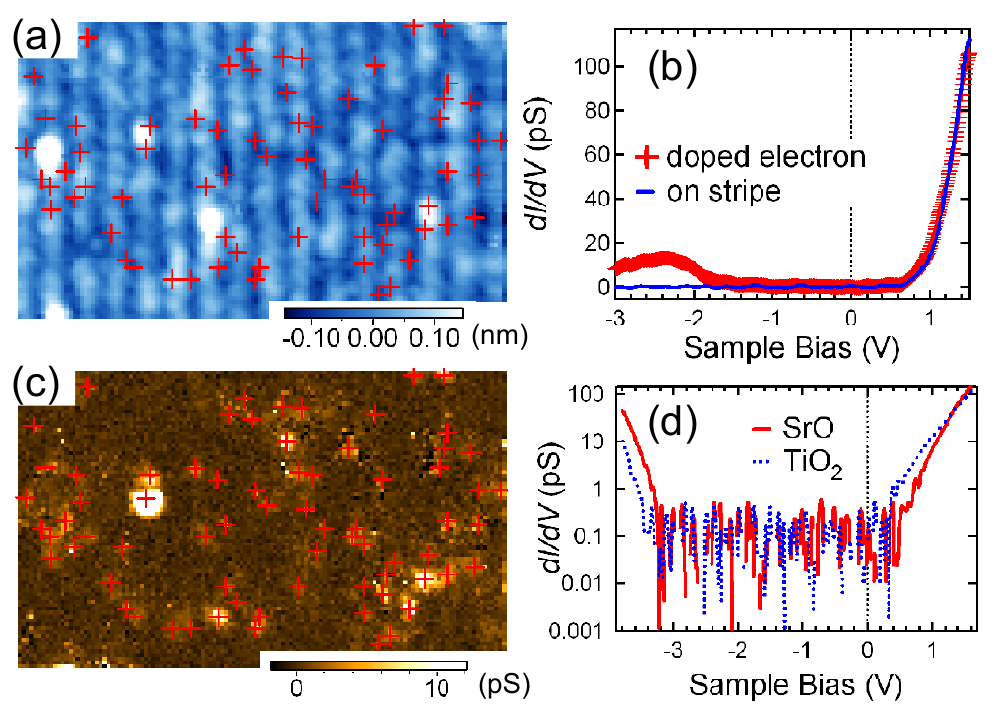}
\caption{(color online)
(a)	STM image of a SrO-terminated SrTiO$_3$(001) thin film surface simultaneously acquired during STS measurement. 
($18 \times 11$ nm$^2$, $V_s = +1.5$ V, $I_t = 30$ pA). 
The marker (+) indicates the location of bright spots observed in the conductance map at $-2.2$ V (c).
(b) d$I$/d$V$ spectrum measured at the bright spot in (a). ($-3$ V $< V_s <$ +1.5 V). 
For comparison, the spectrum at the stripe structure is also shown.    
(c) Conductance map measured at $-2.2$ V. The locations showing finite DOS are indicated by markers. For comparison, these markers are also plotted in (a).   
(d) d$I$/d$V$ spectra measured for the SrO and TiO$_2$ layers. 
Both spectra were measured at locations away from oxygen vacancies. 
}
\label{fig_5}
\end{figure}

Considering the surface preparation was conducted under highly reducing condition at high temperature, 
it is reasonable that the samples used in this study exhibit metallic transport properties \cite{Ohtomo}. 
In such a case, we generally expect to observe a finite DOS at $E_F$ in the tunneling spectrum, 
which is inconsistent with the present results. Previous ARPES measurements for metallic SrTiO$_3$ identified a peak structure at $E_F$ and deep in-gap states, 
but these mainly originated in the bulk states \cite{Aiura,Ishida}.   

In the case of the charge-ordered stripes in La$_{1-x}$Ca$_x$MnO$_3$, 
it is known that large lattice distortions due to the interaction between charge carriers and phonons play a significant role \cite{Mori}. 
Furthermore, a small-polaron ordering was observed along with the charge ordering, 
which indicates a very strong electron-phonon coupling in this system \cite{Li}. 
Polaronic quasiparticle excitations through inelastic tunneling processes were reported 
in manganite systems, with energy gaps evident in the tunneling spectra even in the metallic phase \cite{Ronnow,Seiro}. 

Based on these results, we conclude that the stripe ordering found in this study is exclusively 
induced near the SrO-terminated surface, accompanied by a strong electron-lattice interaction. 
With this scenario, the large energy gaps in the tunneling spectra are likely to be due to the coexistence of the charge ordering and the strong electron-phonon coupling near the surface. 
In addition, the deep in-gap states are possibly associated with doped electrons trapped around atomic defects. 

To date, most efforts to understand various phenomena induced at the oxide interface 
have been implicitly based on the ideal SrO or TiO$_2$ surface termination. 
However, our results have demonstrated the physics in the SrO-terminated SrTiO$_3$ surface, 
which cannot be expected in the ideal SrO termination. 
We believe that an understanding of the local electronic states of the SrTiO$_3$ surface on the atomic scale 
can provide a unique approach for the design of new functional oxide films. 

\section{CONCLUSIONS}

In conclusion, we have performed low-temperature STM measurements to study local electronic states 
of the SrO-terminated SrTiO$_3$(001) surfaces. 
A unidirectional structure was found at low bias voltages, manifested by the stripe charge ordering 
induced in the subsurface TiO$_2$ layer of the SrO-terminated surface. 
In addition, localized in-gap states were observed at deep energies as well as large energy gaps 
in the tunneling spectra. 
These results represent the importance of unidirectional lattice distortion accompanied by a charge ordering 
and a strong electron-lattice interaction in the SrO-terminated surface of SrTiO$_3$.

\begin{acknowledgments}
We would like to thank B. Bryant, I. Hamada, and S. Shiraki for valuable discussions. This study was supported by the World Premier Research Institute Initiative promoted by the Ministry of Education, Culture, Sports, Science, and Technology (MEXT) of Japan and a Grant-in-Aid for Scientific Research from MEXT. 
\end{acknowledgments}



\end{document}